\documentclass[a4paper,10pt,nofootinbib,prd]{revtex4}

\usepackage{graphicx}
\usepackage{amssymb,amsmath}

\begin{document}

\title{Threshold anomalies in Horava-Lifshitz-type theories}

\author{Giovanni AMELINO-CAMELIA}
\affiliation{Dipartimento di Fisica, Universit\`a di Roma ``La Sapienza"\\
and Sez.~Roma1 INFN, P.le A. Moro 2, 00185 Roma, Italy}

\author{Leonardo GUALTIERI}
\affiliation{Dipartimento di Fisica, Universit\`a di Roma ``La Sapienza"\\
and Sez.~Roma1 INFN, P.le A. Moro 2, 00185 Roma, Italy}

\author{Flavio MERCATI}
\affiliation{Dipartimento di Fisica, Universit\`a di Roma ``La Sapienza"\\
and Sez.~Roma1 INFN, P.le A. Moro 2, 00185 Roma, Italy}

\begin{abstract}
Recently the study of threshold kinematic requirements
for particle-production processes has played a very
significant role in the phenomenology of theories
with departures from Poincar\`e symmetry.
We here specialize these threshold studies to
the case of a class of violations of Poincar\`e
symmetry which has been much discussed in the
literature on Horava-Lifshitz scenarios.
These involve modifications of the energy-momentum
(``dispersion'') relation that may be different for
different types of particles, but always involve
even powers of energy-momentum in the correction terms.
We establish the requirements for compatibility with the
observed cosmic-ray spectrum, which is sensitive
to the photopion-production threshold.
We find that the implications for the
electron-positron pair-production threshold
are rather intriguing, in light of some
recent studies of $TeV$ emissions by Blazars.
Our findings should also provide additional motivation for
examining the fate of the law of energy-momentum
conservation in Horava-Lifshitz-type theories.
\end{abstract}

\maketitle

\section{Introduction}
Over the last decade there has been strong
interest
(see, {\it e.g.},
Refs.~\cite{grbgac,astroBiller,kifune,urrutiaPRL,gacQM100,ita,aus,gactp,tedOLDgood,emnPLB2009,gacSMOLINprd,fermiNATURE,atomINTprl})
in the phenomenology
of departures from Poincar\'e symmetry inspired by the literature on the
quantum-gravity problem.
The recent interest in the Horava-Lifshitz scenario
(see, {\it e.g.}, Refs.~\cite{H09,Horavabig,CNPS09})
offers yet another
opportunity to make use of this line of investigation,
since indeed this scenario is motivated by the
study of the quantum-gravity problem and is based on a mechanism
that produces violations of Poincar\'e symmetries.
While a detailed understanding of the nature of these Horava-Lifshitz-type
Poincar\'e-symmetry violations has still not been reached, mostly because of
challenges from a renormalization-group perspective~\cite{IRS09,OR09},
some consensus appears to be emerging at least on some features that can be
used, as already stressed by other authors~\cite{Horavaphen,Horavaphen2},
for a preliminary phenomenological analysis.

We here intend to expose some opportunities that are found in the
study of ``threshold anomalies"~\cite{gactp}, {\it i.e.} the study of the
implications of violations of Poincar\'e symmetry for the kinematic conditions at the
threshold for some particle-creation interactions. The main feature
we shall use in the analysis is the presence of modifications
of the energy/momentum (dispersion) relation of the type
$$E^2 = m_i^2 + p^2 + \lambda_i^{(2)} p^4 + \lambda_i^{(4)} p^6 $$
where the dimensionful parameters $\lambda_i^{(n)}$ carry an
 index $i$ which denotes a possible ``non-universality" of the effects
(effects that have different magnitude for different particles)
and an index $(n)$ which simply refers to the number of length
dimensions ({\it e.g.}, $dim[\lambda_i^{(2)}]= [l^2]$).

The fact that this type of dispersion relations may emerge in
the Horava-Lifshitz scenario has been argued by several
authors (see, {\it e.g.}, Refs.~\cite{Horavaphen,Horavaphen2}).
But our analysis might be valuable
from a broader perspective, with or without the support of
the Horava-Lifshitz literature. Previous studies of threshold anomalies
(see, {\it e.g.}, Refs.~\cite{kifune,gacQM100,ita,aus,gactp,tedOLDgood})
focused on the possibility that dimensionful parameters
such as $\lambda_i^{(2)}$ and $\lambda_i^{(4)}$ be set by the Planck
scale (so that, {\it e.g.}, $\lambda_i^{(2)} \sim 1/E_{planck}^2 \simeq 10^{-56}~eV^{-2}$),
but this might provide an incomplete characterization of the possibilities
this phenomenology offers. In this respect the picture which is emerging
from the Horava-Lifshitz scenario, in which $\lambda_i^{(2)}$, $\lambda_i^{(4)}$
are not directly linked to the Planck scale, provides of course explicit motivation.
Another limitation of previous phenomenological analyses of threshold anomalies
concerns the handling of the energy-momentum conservation law. In other
frameworks with violations of Poincar\'e symmetry only the case of unmodified
energy-momentum conservation was considered.
Modifications of energy-momentum conservation were considered in several studies
but only when attempting to ultimately restore (at least deformed)
relativistic invariance, in the sense of the ``Doubly Special Relativity"
proposal~\cite{gacdsr1,gacdsr2,leejoaoPRDdsr}.
Also concerning the possibility of violations of energy-momentum conservation
in Poincar\'e-violation scenarios the present understanding of the
Horava-Lifshitz framework exposes the need for more general analyses,
since it involves modified dispersion relations within a framework
that also appears to involve violations of translational
symmetries~\cite{CNPS09,M09c}, which may well produce
violations of energy-momentum conservation.

In the next section we consider threshold anomalies for the process
of electron-positron pair production in photon-photon collisions,
and find that our Horava-Lifshitz-inspired analysis leads to a picture
that could produce an increase in our expectations for the spectrum
of multi-$TeV$ photons to be observed from certain Blazars.
Since it has been argued~\cite{aus,piranIRABS} that indeed the abundance
of multi-$TeV$ photons observed from certain Blazars is unexpectedly high this
may be a valuable opportunity for
Horava-Lifshitz phenomenology.
Section 3 discusses an analogous threshold anomaly for photopion production,
which is relevant for the observations of ultra-high-energy cosmic rays,
and provides the basis for additional insight on what type
of Horava-Lifshitz-inspired models could provide the most fruitful
phenomenology, while preserving consistence with available
experimental data.
While Section 2 and 3 take as working assumption the fact
that possible effects of modification of the law
of energy-momentum conservation can be neglected at the level of our leading-order
analysis, in Section 4 we discuss the differences in the description of threshold
anomalies that would instead arise if leading-order effects of modification
of energy-momentum conservation were to be found. The main objective
of Section 4 is therefore the one of highlighting the significance
for phenomenology of the analysis of violations of translational symmetry
in the Horava-Lifshitz scenario, which unfortunately has so far not attracted
much attention.
Section~5 offers some closing remarks.

\section{Pair-production  threshold anomalies}
The study of the threshold kinematic requirements
for the pair-production process, $\gamma \gamma \rightarrow e^+ e^-$,
has important implications for
 the opacity
of the Universe to photons, which in turn can be indirectly studied observationally.
In previous quantum-gravity-motivated studies~\cite{ita,aus,gactp,tedOLDgood}
of anomalies
for the pair-production threshold it was already observed that
violations of Poincar\'e symmetry
can be particularly significant for
the study of absorption of  multi-$TeV$ photons
(photons with energies between a few and, say, $30~TeV$)
by the infrared diffuse extragalactic background.
In this section we intend to specialize this observation to the
case of the
Horava-Lifshitz-inspired phenomenological framework described in
our introductory remarks, centered on a modification of the dispersion relation.

The fact that we plan to obtain results relevant for collisions
between a multi-$TeV$ photon
and a photon in the infrared diffuse extragalactic background
invites us to consider the case of a collision in which
one of the photons is hard, with energy-momentum $E,P$
such that $E \gg m_e$ (denoting with $m_e$ the electron mass), whereas
the other photon is soft, with energy-momentum $\epsilon,p$
such that $\epsilon \ll m_e$.
Of course, for fixed value of the soft-photon energy $\epsilon$
(representative of photons in the infrared diffuse background)
the production of an electron-positron pair is possible only
for values of the hard-photon energy $E$ greater than a certain minimum
(threshold) value, which we can denote with $E_\epsilon^*$.
Within ordinary Poincar\'e covariant kinematics
one easily finds that the threshold requirement
is $E > E_\epsilon^* = m_e^2 /\epsilon$,
but it is known~\cite{kifune,gacQM100,ita,aus,gactp,tedOLDgood}
that this
result can be affected rather sizably even by
small departures from Poincar\'e symmetry.

In order to establish the size of the threshold anomaly for the
case of our Horava-Lifshitz-inspired framework we shall of course make use of the modified
dispersion relation already discussed in the introductory remarks.
For the hard photon we therefore have
\begin{equation}
E \simeq P + \frac{1}{2} \lambda_\gamma^{(n)} ~ P^{n+1} ~, \label{DisprelPhot}
\end{equation}
where $n$ can be 2 or 4.
We are only aiming for a description of the dominant correction
to the threshold requirement, so we will only consider $n=2$ whenever
its implications are not negligible with respect to the ones of the $n=4$ terms,
and in turn consider exclusively the $n=4$ terms if the contributions
from terms with $n=2$ can be neglected.
We assume of course that the $\lambda_i^{(n)}$ are all small, and in particular
in the analysis of the pair-production threshold
we assume $\lambda_{\gamma,e}^{(2)} \ll (100~TeV)^{-2}$
and $\lambda_{\gamma,e}^{(4)} \ll (100~TeV)^{-4}$.
On dimensional ground one might
guess $\lambda_i^{(4)} \sim \left(\lambda_i^{(2)}\right)^2$, and
whenever  $\lambda_i^{(4)} \lesssim \left(\lambda_i^{(2)}\right)^2$
the effects
of the $\lambda_i^{(4)}$ parameters can be neglected at leading order.
As already observed in Ref.~\cite{Horavaphen}, only
in the case $\lambda_i^{(4)} \gg \left(\lambda_i^{(2)}\right)^2$
the $\lambda_i^{(4)}$ parameters produce the dominant effects.

Consistently with the scopes of our leading-order analysis we can neglect
the modification of the soft-photon dispersion relation,
\begin{equation}
\epsilon \simeq p + \frac{1}{2} \lambda_\gamma^{(n)} ~ p^{n+1}\simeq p  ~, \label{DisprelPhotSOFT}
\end{equation}
since we are interested in the case of $p \ll P$, which
of course implies that $\lambda_\gamma^{(n)} p^{n+1} \ll \lambda_\gamma^{(n)} P^{n+1}$.

For the outgoing electron (positron)
we introduce the notation $E_-$ ($E_+$) for its energy and $p_-$ ($p_+$)
for its spatial momentum, so that
\begin{equation}
E_\pm \simeq p_\pm + \frac{m_e^2}{2 p_\pm} + \frac{1}{2} \lambda_e^{(n)} ~ p_\pm^{n+1}~, \label{DisprelElec}
\end{equation}
where we also used the fact that the electron-positron pairs produced at threshold in
collisions between a
multi-$TeV$ photon
and a photon in the infrared diffuse extragalactic background
are inevitably ultra-relativistic ($ p_\pm \gg m_e $).

The kinematic requirements at threshold are the ones that require the minimum
energies for the process to occur and as a result the process at threshold
inevitably is a head-on collision~\cite{gactp} (collisions that are not head-on
always ``cost" more energy, which ``pays for" the additional components of momentum).
This simplifies the analysis since for the purpose of establishing the
threshold requirements one can exploit the fact that the whole process
occurs along one spatial direction. We can therefore efficaciously reason
in terms of the modulus of the spatial momenta, and write energy-momentum
conservation as follows:
\begin{equation}
\left\{
\begin{array}{l}
E + \epsilon = E_+ + E_-
\\
P - p = p_+ + p_-
\end{array}
\right. \label{ConservationLaw}
\end{equation}
Using the dispersion relations (\ref{DisprelPhot})
and (\ref{DisprelElec}) in the equation of conservation of energy
one finds
\begin{equation}
P +\frac{1}{2} \lambda_\gamma^{(n)} ~ P^{n+1} + \epsilon
= p_+ + p_- + \frac{m_e^2}{2 p_+} + \frac{m_e^2}{2 p_-}
+ \frac{1}{2} \lambda_e^{(n)} ( p_+^{n+1} + p_-^{n+1}) ~,
\end{equation}
which can then be cast in the form
\begin{equation}
\frac{1}{2} \lambda_\gamma^{(n)} ~ P^{n+1} + 2 \epsilon
=  \frac{m_e^2}{2 p_+} + \frac{m_e^2}{2 p_-} +\frac{1}{2} \lambda_e^{(n)} ( p_+^{n+1} + p_-^{n+1})
\label{nMINUS1}
\end{equation}
using the equation of conservation of spatial momentum and the fact that $\epsilon \simeq p$.

Next we observe that
at zero-th order in $\lambda^{(n)}_\gamma$, $\lambda^{(n)}_e$
({\it i.e.} in the standard Poincar\'e covariant derivation)
one obtains from these equations $p_+=p_- \simeq P/2$.
This can be exploited in our first-order derivation by allowing us
to observe that
\begin{equation}
\frac{m_e^2}{p_+}+\frac{m_e^2}{p_-} \simeq \frac{4 m_e^2}{P}
~,
\qquad \lambda^{(n)}_e p_\pm^{n+1} = \lambda^{(n)}_e \left(\frac{P}{2}\right)^{n+1}
~,
\label{zerothbalance}
\end{equation}
neglecting terms on magnitude not greater than $\lambda^{(n)} P^{n+1} m_e^2/P^2$
which of course can be neglected in our derivation focusing on the
dominant $\lambda^{(n)} P^{n+1}$ corrections ($m_e/P$ is indeed very small for
the collisions we intend to investigate).
Using (\ref{zerothbalance}) one obtains from (\ref{nMINUS1}) the following
result
\begin{equation}
P +  \lambda_\gamma^{(n)} \frac{P^{n+2}}{4\epsilon} \simeq \frac{m_e^2}{\epsilon}
+ \lambda_e^{(n)} ~ \frac{P^{n+2} }{2^{n+2} \epsilon} ~.
\end{equation}
And in turn this allows us to conclude that, for fixed soft-photon energy $\epsilon$, the pair-production
process is possible within our Horava-Lifshitz-inspired framework
only when $E>E_\epsilon^*$, where the threshold energy $E_\epsilon^{*}$
is solution of the equation
\begin{equation}
 E_\epsilon^* +\left(\lambda_\gamma^{(n)}
 - \frac{\lambda_e^{(n)}}{2^n} \right)  \frac{(E_\epsilon^*)^{n+2}}{4\epsilon}
  \simeq \frac{m_e^2}{\epsilon}
\end{equation}
The standard Poincar\'e covariant result  $E_\epsilon^* = m_e^2 /\epsilon$
is of course recovered in the
limit $\lambda_\gamma^{(n)},\lambda_e^{(n)} \rightarrow 0$.
For $\lambda_\gamma^{(n)} > \lambda_e^{(n)}/{2^n}$ one finds
lower values of $E_\epsilon^*$,
while for $\lambda_\gamma^{(n)} < \lambda_e^{(n)}/{2^n}$
one obtains
 values of $E_\epsilon^*$ that are greater than $m_e^2 /\epsilon$.

Analogous relations among parameters of schemes with particle-dependent (non-universal)
modifications of the dispersion relations have already been derived
(see, {\it e.g.}, Ref.~\cite{tedOLDgood}),
but typically assuming
symmetry-breaking scales of the order of the Planck scale ($\sim 10^{28}~eV$)
and different dependence on energy.
Taking $E \sim 10~TeV$ and $\epsilon \sim 0.04~eV$ one easily
verifies that, in order for our Horava-Lifshitz-inspired framework to have observably large
implications in this pair-production analysis,
the scales $\lambda_\gamma^{(n)}$ and/or $\lambda_e^{(n)}$ should not be set by the Planck
 scale but by a much lower scale. For example in the case $n=2$
 one would need $|\lambda_{\gamma ,e}^{(2)}| \gtrsim (10^{20}~eV)^{-2}$.

Preliminary indications on whether values higher or lower
than $m_e^2 /\epsilon$ could be favored experimentally
can be obtained using data on the opacity of the Universe
for multi-$TeV$ photons.
A high energy photon propagating in
the intergalactic space can indeed interact with
photons in the infrared diffuse extragalactic background,
producing an electron-positron
pair.
The mean free path of 10 TeV photons depends on the spectrum of the
infrared background photons in the range from $\simeq 0.03~eV$ to $\simeq 0.08~eV$,
with particularly strong dependence on the spectrum around $0.04~eV$. And these
estimates scale linearly with the (inverse of)
the energy of the incoming hard photon.
Unfortunately, it is difficult to
determine the  infrared diffuse extragalactic background, since
direct measurements are problematic, owing to the presence of the
bright Galactic and Solar System foregrounds~\cite{piranIRABS}. Still it is noteworthy
that in recent years there have been several reports (see, {\it e.g.},
Refs.~\cite{piranIRABS,Aharonian,magicFIRB} and references therein) of spectra of some
observed blazars that appear to be harder than anticipated on the basis
of the expected infrared-background absorption.
One could therefore tentatively argue that the case of values of the
pair-production threshold that are somewhat higher than $m_e^2 /\epsilon$,
{\it i.e.} the
case $\lambda_\gamma^{(n)} \lesssim \lambda_e^{(n)}/{2^n}$,
finds some encouragement in the, however preliminary, observational situation.
But this possibility must be contemplated very cautiously since the presence
of anomalies is in no way necessary~\cite{steckerSIAMOFELICI}.
The observational situation does establish more robustly that
values of the
pair-production threshold lower than $m_e^2 /\epsilon$
are objectively disfavored \cite{gactp}, so that
Horava-Lifshitz scenarios with $\lambda_\gamma^{(n)} > \lambda_e^{(n)}/{2^n}$
(and $|\lambda_{\gamma ,e}^{(2)}| \gtrsim (10^{20}~eV)^{-2}$)
appear to be excluded.

\section{Photopion-production threshold anomalies}

In the preceding subsection we discussed the implications of
Horava-Lifshitz deformed dispersion relations for the
process $\gamma \gamma \rightarrow e^+ e^-$, but of course this is not the only process
in which deformations to dispersion relations can produce significant threshold anomalies.
In particular,  there has been strong
interest~\cite{kifune,ita,gactp,aus,tedOLDgood}
in the analysis of the threshold requirements
for the ``photopion production" process, $p \gamma \rightarrow p \pi$,
and their relevance for the observed high-energy portion of the
cosmic-ray spectrum.

The analysis of the photopion-production threshold is of course
completely analogous to the one of the pair-production threshold,
but it is slightly more tedious:
in the case of $\gamma \gamma \rightarrow e^+ e^-$
the calculations are simplified
by the fact that both outgoing particles have the same mass
and both incoming particles are massless,
whereas for the threshold conditions for the
photopion-production process one needs to
handle the kinematics for a head-on
collision between a soft photon of energy $\epsilon$
and a high-energy particle of mass $m_p$ and momentum $P_p$
producing  two (outgoing) particles with masses $m_p$, $m_\pi$ and
momenta $P'_{p}$, $P_\pi$.
Since however these additional complications pose no conceptual and no significant technical
challenges (and a dedicated derivation of the photopion-production threshold with
Poincar\'e-symmetry violations is given in Ref.~\cite{gactp}) we shall here
just note the final result for
the threshold condition in our Horava-Lifshitz-inspired framework:
\begin{equation}
E_\epsilon^* +
\frac{(E_\epsilon^* )^{2+n} }{ 4 \epsilon} \left[ \lambda_p^{(n)}
- \lambda_p^{(n)} \left(\frac{m_p}{m_p + m_\pi} \right)^{n+1}
- \lambda_{\pi}^{(n)} \left(\frac{m_\pi}{ m_p + m_\pi} \right)^{n+1}
\right]
\simeq \frac{(m_p + m_\pi)^2 - m_p^2}{4 \epsilon}
\label{lithresh2}
\end{equation}
(neglecting of course all terms suppressed
by both the smallness of $\lambda_i^{(n)}$ and the smallness of $\epsilon$
and/or $m_{p,\pi}$).

Introducing the notation $\mu_p \equiv m_p/(m_p + m_\pi) \simeq 0.9$
and $\mu_\pi \equiv m_\pi/(m_p + m_\pi) \simeq 0.1$
one therefore concludes that,
for fixed soft-photon energy $\epsilon$,
when
$\lambda_p^{(n)} (1 -\mu_p^{n+1}) > \lambda_\pi^{(n)} \mu_\pi^{n+1}$
the energy of the incoming proton
required at threshold
for photopion production is shifted toward lower values (in comparison to
the standard case $\lambda_p^{(n)}=\lambda_{\pi}^{(n)}=0$),
whereas when
$\lambda_p^{(n)} (1 -\mu_p^{n+1}) < \lambda_\pi^{(n)} \mu_\pi^{n+1}$
this threshold energy is shifted toward higher values.

An exciting aspect of these threshold analyses for photopion production and
the cosmic-ray spectrum is that they in principle provide access to scales
of violation of Poincar\'e symmetry that are extremely high.
For example, from (\ref{lithresh2}) it is easy to infer
that detailed studies of the cosmic-ray spectrum
at energies $\gtrsim 10^{19}~eV$
could allow us to probe values of $\lambda_p^{(2)}$
and $\lambda_\pi^{(2)}$ such that $|\lambda_{p,\pi}^{(2)}| \gtrsim (10^{30}~eV)^{-2}$.

The feature of the cosmic-ray spectrum that can be most valuable
from this perspective is associated
with the Greisen-Zatsepin-Kuzmin (GZK) cutoff,
which is essentially obtained as the
threshold energy ($\sim 5 \cdot 10^{19}~eV$)
for cosmic-ray protons to produce pions in
collisions with CMBR photons.
The observational determination of the cosmic-ray spectrum has recently improved rather
significantly as a result of observations conducted with the Pierre Auger cosmic-ray
observatory~\cite{augersources}. There is no evidence of any
shift of the GZK threshold within the accuracy so far achieved in determining
the cosmic-ray spectrum,
but the most promising outlook from the perspective of possible Poincar\'e violations
is the one discussed in Ref.~\cite{steckerAUGERnew},
which, within the framework here considered, would
require $\lambda_p^{(n)} (1 -\mu_p^{n+1}) < \lambda_\pi^{(n)} \mu_\pi^{n+1}$.
This scenario of Ref.~\cite{steckerAUGERnew} ensures consistency with available
cosmic-ray-spectrum data and predicts a sort of ``recovery" \cite{steckerAUGERnew}
of the spectrum at energies not much higher than the GZK scale.
The prospects are therefore rather intriguing since
a better determination of the beyond-GZK portion of the spectrum
appears to be within the reach of studies planned for the Pierre Auger observatory.

\section{A possible role for modifications of energy-momentum conservation}
The results we derived so far assume that in the Horava-Lifshitz scenario
there are no modifications of the law of energy-momentum conservation that
could be large enough to affect the threshold requirements at the
leading $\lambda^{(n)} P^{n+1}$ order.  Previous studies~\cite{kifune,gacQM100,ita,aus,gactp,tedOLDgood}
of the phenomenology of threshold anomalies due to violations of Poincar\'e symmetry
mainly focused on the possibility of new physics affecting exclusively the Lorentz sector,
so that the law of energy-momentum conservation would be unaffected.
However, in the Horava-Lifshitz scenario the four-dimensional
diffeomorphism invariance is broken down to foliation-preserving diffeomorphisms
\begin{equation}
\delta x^i=\zeta^i(t,{\bf x})\,,~~~~~\delta t=f(t)\,,
\end{equation}
a subgroup which preserves the foliation structure
of space-like slices. Therefore, as pointed out in
several studies (see, {\it e.g.}, Refs.~\cite{CNPS09,M09c}),
local energy-momentum conservation is restricted to the spatial
components.
In a locally inertial frame, the theory is invariant under space translations
but not under time translations, so that in principle energy might not be conserved.

Presently the literature still does not provide any guidance
on the magnitude of the violations (if any) of energy conservation in particle-physics
processes within the Horava-Lifshitz framework. But it is important
for us to stress that our results could be significantly changed
if these violations happen to be relevant, also hoping that this observation might
motivate a more intense phase of study by the community of the
issue of energy conservation in the Horava-Lifshitz framework.

For our exclusively illustrative purposes here
it is sufficient to make a simple ansatz for a modified
law of energy-momentum conservation, applicable to the case of electron-positron
pair production in collisions between two photons:
\begin{equation}
\left\{
\begin{array}{l}
E + \epsilon - \Delta^{\!\!(2)}  \left( E \epsilon^2 + E^2 \epsilon \right)
= E_+ + E_- - \Delta^{\!\!(2)}  \left( E_+^2 E_- + E_+ E^2_- \right)
\\
P - p = p_+ + p_-
\end{array}
\right.
\end{equation}
where $\Delta^{\!\!(2)} $ is a parameter with length-squared dimensions.
We use this recipe to obtain a rough estimate of the size of
the threshold-anomaly effects that could be induced by violations of energy conservation
with $P^3$ behaviour.
And we shall be satisfied showing the implications of the parameter $\Delta^{\!\!(2)} $
for the case of the pair-production threshold, focusing on the dispersion-relation
parameters with $n=2$, $\lambda_\gamma^{(2)}$ and $\lambda_e^{(2)}$.
Adopting the $\Delta^{\!\!(2)} $-deformed energy-momentum conservation, the
derivation of the pair-production threshold requirement (which of course
once again follows exactly
the same steps described in Section~2) leads to the result
\begin{equation}
 E_\epsilon^* +\left(\frac{\Delta^{\!\!(2)} }{2} + \lambda_\gamma^{(2)}
 - \frac{\lambda_e^{(2)}}{4} \right)  \frac{(E_\epsilon^*)^{4}}{4\epsilon}
  \simeq \frac{m_e^2}{\epsilon} ~.
\end{equation}
This shows that modifications of the law of energy-momentum conservation of magnitude
comparable to the one we illustratively considered, and parametrized
with $\Delta^{\!\!(2)} $, could affect the result for the threshold at the same
level as the $\lambda_i^{(2)}$ parameters of modification of the dispersion relation.
In principle one could even have cases in which the modification of the
dispersion relation and the modification of the law of energy-momentum
conservation balance each other
($\Delta^{\!\!(2)}  =\lambda_e^{(2)}/2 -2 \lambda_\gamma^{(2)}$)
giving the net result of no leading-order correction to the threshold requirements.
Such a cancellation is actually expected~\cite{gacdsr1,dsrphen} in frameworks based on the concept
of ``Doubly Special Relativity"~\cite{gacdsr1,gacdsr2,leejoaoPRDdsr},
where one could accommodate modifications of the dispersion relation within
a model which is still fully relativistic, but relativistic in a deformed sense
(with two nontrivial relativistic invariants, a speed scale and a length scale,
rather than one). But such a cancellation is not to be expected~\cite{dsrphen} in frameworks
in which instead Poincar\'e symmetry is genuinely broken (rather than deformed)
as appears to be the case of the Horava-Lifshitz framework. So, while we cannot exclude that
investigations of the fate of the relevant diffeomorphism-invariance issues
may lead to a reassessment of the quantitative aspects (magnitude)
of the threshold anomalies we here considered,
we do expect these threshold anomalies to be a genuine characteristic
of the  Horava-Lifshitz framework.

\section{Closing remarks}
In spite of a vigorous effort, composed of a large number of dedicated studies in just a short time, the understanding of the physics of the Horava-Lifshitz scenario appears to be still far from taking final shape. There is however growing consensus on some aspects, and particularly on the presence~\cite{Horavaphen,Horavaphen2} of modifications of the dispersion relation of the type we here studied.  The threshold anomalies we analyzed represent challenges and opportunities which may provide guidance, and perhaps even encouragement, for further studies of the framework.

From a phenomenology perspective interest in this scenario can originate
from the rather natural emergence of ``non-universal effects" (different magnitude for
different type of particles), but in ways that one can imagine to become predictive
at a later more mature stage of investigation. Particularly interesting from our perspective
is the possibility that one might find that the implications of the
Horava-Lifshitz scenario are different for particles of different spin, since our analysis
involved particles with spin 1, 1/2 and 0 ({\it i.e.} $\gamma$, $e^{\pm}$, $p$, $\pi$).
For example, the most intriguing
aspect of our analysis concerns the pair-production threshold, where the observations
appear to invite (however prudently) consideration of the possibility of new fundamental physics.
The requirement we obtained, $\lambda_\gamma^{(n)} \lesssim \lambda_e^{(n)}/{2^n}$,
would carry little significance if one ended up introducing it by hand
in the  Horava-Lifshitz scenario, since it would then amount to
a standard observationally-imposed constraint on a potentially rather large parameter
space. But the present limited understanding of the framework, particularly
for what concerns issues connected with the renormalization group~\cite{IRS09},
appears to leave open the possibility that such a condition be derived
as an inevitable feature of the Horava-Lifshitz setup. In that case
the evaluation of compellingness of the proposal should clearly take into account
the type of phenomenological implications that we here focused on.

Of similar nature is our contribution on the points concerning the law of energy-momentum
conservation. In that respect the most interesting aspect from the phenomenology perspective
originates from the fact that the Horava-Lifshitz scenario might host both
modifications of the dispersion relation, of a type that is not
too different from the ones already considered in other Poincar\'e-violation
scenarios, and modifications of the law of energy-momentum conservation,
which is instead a possibility that had been mostly neglected in
previous studies of Poincar\'e-violation scenarios.
We observed here that there could be a strong dependence
of a meaningful observable aspect (our threshold anomalies)
on possible violations of the law of energy-momentum conservation,
also hoping to provide motivation for an increased effort of investigation
of the fate of translational symmetries in the  Horava-Lifshitz scenario.
In spite of the large number of studies devoted to this proposal, only very
few authors appear to have considered the implications for translational symmetries,
which instead, in ways that our analysis renders more tangible, will
probably play a key role in assessing the compellingness of the physical
picture produced by the  Horava-Lifshitz scenario.

\begin{acknowledgments}
This work was supported in part by a grant from the Ateneo Federato della Scienza
e Tecnologia "Nuova iniziativa di ricerca di ateneo federato AST-2008".  G. A.-C.
also acknowledges support by grant RFP2-08-02 from The Foundational Questions
Institute (fqxi.org).  L.G. has been supported in part by the grant PTDC/FIS/098025/2008.
\end{acknowledgments}

\end{document}